\documentstyle[amssymb,preprint,aps]{revtex}

\begin{document}
\title{{\bf Search for Spontaneous Nucleation of Magnetic Flux During Rapid Cooling
of YBa}$_{2}${\bf Cu}$_{3}${\bf O}$_{7-\delta }${\bf \ Films Through }$T_{c}$}
\author{Raz Carmi and Emil Polturak}
\address{Physics Department, Technion-Israel Institute of Technology, Haifa 32000,\\
Israel.}
\maketitle

\begin{abstract}
We describe an experimental search for spontaneous formation of flux lines
during a rapid quench of thin YBa$_{2}$Cu$_{3}$O$_{7-\delta }$\ films
through $T_{c}$. This effect is expected according to the Kibble-Zurek
mechanism of a creation of topological defects of the order parameter during
a symmetry breaking phase transition. Spontaneously formed vortices were
previously observed in superfluid $^{3}$He, while a similar experiment in
superfluid $^{4}$He gave negative results. Using a high $T_{c}$ SQUID, we
measured both the magnetic flux in the sample during a quench with a
sensitivity of 20 $\phi _{0}/$cm$^{2}$, and the field noise which one would
expect from flux lines pinned in the film. The sensitivity was sufficient to
detect spontaneous flux at a level corresponding to 10$^{-3}$ of the
prediction. Within our resolution, we saw no evidence for this effect.
\end{abstract}

 \vskip 0.2cm

\newpage

\section{Introduction}

If a system undergoing a phase transition into an ordered state is quenched
through the phase transition fast enough, topological defects can be created
due to the evolution of uncorrelated regions of the newly formed phase,
having different values of the order parameter. The defects appear at the
boundary separating several coalesced regions of this kind. Such a scenario
was first proposed by Kibble\cite{Kibble} in the context of the Grand
Unified Theory, in order to describe the symmetry-breaking phase transition
in the early universe. Zurek\cite{Zurek} developed this idea to predict the
initial density of defects created during the phase transition and suggested
specific experiments on condensed matter systems to test this scenario.
Because of the generality of the theory of phase transitions, topological
defect formation should occur in every physical system having a relevant
symmetry breaking of the order parameter during the transition. The main
arguments{\bf \ }used by Zurek rely on the critical slowing down of
fluctuations and the divergence of the coherence length near a second order
phase-transition. Both quantities are influenced dynamically by the
characteristic time required to complete the transition. In the last few
years, experiments were carried out on several systems: first, nematic
liquid crystals undergoing an isotropic-nematic transition\cite
{Chuang,Bowick} (In this case the topological defects are disclinations).
Another system is liquid $^{4}$He crossing the $\lambda $ transition as a
result of rapid depressurization\cite{Hendry,Dodd}; in this case the
topological defects are quantized vortex lines. A third system is liquid $%
^{3}$He undergoing its superfluid phase transition\cite{Bauerle,Ruutu}. In
this experiment, the quantized vortices are formed during a thermal quench
induced following an exothermic neutron-induced nuclear reaction. The
experiments on liquid crystals and $^{3}$He gave results consistent with
Zurek's prediction. In contrast, after early claims{\bf \ }of observing this
effect in liquid $^{4}$He\cite{Hendry}, a recent improved experiment showed
no spontaneous nucleation of vortices within the experimental resolution\cite
{Dodd}. Here, we report an analogous experiment with a superconductor; the
topological defects are quantized flux lines. Specifically, the experiment
aims to observe spontaneous flux lines generated during thermal quench of YBa%
$_{2}$Cu$_{3}$O$_{7-\delta }$ (YBCO) thin films through the normal to
superconductor transition. The additional importance of this experiment is
to test the ''cosmological'' scenario of Kibble and Zurek\cite{Kibble,Zurek}
in a system with a local gauge symmetry, where the theory is less clear then
in a systems having a global gauge (e.g. $^{4}$He and $^{3}$He), due to the
evolving gauge field ({\bf B}) during the transition\cite{Rudaz,Kibble2}.
Within our resolution, we found no spontaneous vortex formation down to a
level of 10$^{-3}$ of the predictions, which is in variance with the
original theory of Zurek.\ 

\section{The Kibble-Zurek mechanism in a superconductor}

Flux lines may become created spontaneously in type II superconductors
during a rapid quench through $T_{c}$. The most practical way is a thermal
quench (which is more reliable from pressure quench in these materials,
since the pressure dependence of $T_{c}$ is rather weak). Low temperature
superconductor (LTS) have a second order phase transition which is well
described by the Landau-Ginzburg theory and a rather small critical region,
thus, the anticipated initial flux line density should be well predicted by
Zurek's theory\cite{Zurek}. According to these predictions, the {\it initial}
vortex density after the quench should be:

\begin{equation}
n_{i}\approx \frac{1}{\xi _{0}^{2}}(\tau _{0}/\tau _{Q})^{p}  \label{indens}
\end{equation}

Here, $\xi _{0}$ and $\tau _{0}$ are the coherence length and the relaxation
time of the order parameter at $T=0$ respectively. The typical quench time, $%
\tau _{Q}$, is defined as: $\tau _{Q}=\frac{1}{(d\epsilon /dt)_{\epsilon =0}}
$ , $\varepsilon $ being the reduced temperature. The exponent $p$ is
related to the critical exponents of the coherence length and the relaxation
time as: $p=(\frac{2\nu }{\mu +1})$, where: $\xi =\xi _{0}\left| \varepsilon
\right| ^{-\nu }$, $\tau =\tau _{0}\left| \varepsilon \right| ^{-\mu }$.
From an experimental point of view, the problem with LTS is that their
coherence lengths are of the order of 100nm and therefore the flux line
density predicted by Eq. (\ref{indens}) is quite small. A more favorable
situation exist in high temperature superconductors (HTS) which have a much
shorter coherence length ($\xi _{0}\simeq 2$nm). The phase transition of HTS
is closer to the 3D-XY model, and has a wider critical region than LTS, but
as both the coherence length and the relaxation time diverge at $T_{c}$,
that should be sufficient for spontaneous vortex formation (provided the
quench is fast enough). Another important aspect unique to a superconductor
is flux pinning, which can cause a significant reduction of the mutual
annihilation of vortices and antivortices generated during the quench.
Pinning can also prevent the flux from being expelled out of the film.
Finally, within the last few years it was found that the order parameter in
HTS has a predominant $d$-wave symmetry, while in metal-LTS it is always $s$%
-wave. This type of paring can lead to spontaneous flux generation in some
special configurations (ref. \cite{harlingen} and other ref. therein), but
in a homogenous material it should not affect the Kibble-Zurek scenario. The
values pertinent to YBCO are $\xi _{0}\simeq 1.6$ nm , $\tau _{0}\simeq
5\cdot 10^{-12}$ sec\cite{Han}. In our experimental setup, described below,
we achieve a 20K/sec cooling rate, giving $\tau _{Q}\simeq 5\sec .$

Because the phase transition in YBCO is not a pure Ginzburg-Landau type,
closer to the 3D-XY model, the value of the exponent $p$ is deduced from
experiments that measured the exponents $\nu $ and $\mu $ \cite
{Overend,Kamal,Booth,Roberts}. We chose average values: $\nu \simeq 0.67$, $%
\mu \simeq 3.4$ (note that this is an unusual scaling), and then $p\simeq 0.3
$. Due to the short coherence length, the predicted {\it initial} flux line
density generated in the film by a thermal quench is very large: $n_{i}\sim
10^{10}$ cm$^{-2}$ (vortices and antivortices). An important quantity is the
reduced temperature $\widehat{\varepsilon }$ at which the system returns to
the usual critical behavior during the quench, after an initial period
during which it is out of equilibrium. Spontaneous flux nucleation takes
place around this temperature, and thus determines the initial vortex
density. According to the theory: $\widehat{\varepsilon }=(\tau _{0}/\tau
_{Q})^{1/(\mu +1)}$. In our experiment $\widehat{\varepsilon }$ is of the
same order of magnitude as the width of the regime in which strong
fluctuations can erase flux lines $(\varepsilon _{f}\sim 10^{-3})$, as
deduced from measurements of flux line noise in thermal equilibrium\cite
{Ferrari}. However, as soon as the temperature decreases and the flux lines
become pinned, one can hope to preserve a significant fraction of the
initial flux line density. In our system, the situation is more favorable in
this respect than in superfluid $^{4}$He, since there the regime of strong
fluctuations is wider. Although the value of $\widehat{\varepsilon }$ is not
necessarily determined by $\varepsilon _{G}$ from the Ginzburg criterion\cite
{Gill} (in $^{4}$He: $\varepsilon _{G}\sim 0.2$) it is not reasonable that
it can be smaller by more than a few orders of magnitude. In the recent
experiment\cite{Dodd} $\widehat{\varepsilon }\sim 10^{-5}$ {\bf .} Moreover,
there is no pinning of vortices in bulk $^{4}$He which can help to preserve
a significant fraction of these vortices. The freeze-out coherence length $%
\widehat{\xi }$, corresponding to $\widehat{\varepsilon }$, is $\widehat{\xi 
}=\xi _{0}(\tau _{Q}/\tau _{0})^{\nu /(\mu +1)}$. We get $\widehat{\xi }\sim
0.1\mu $m, of the same order of magnitude as the thickness of the films
(50-300nm). Therefore, the initial vortex array is two dimensional (but the
physical system is $3D$). In our experiment we can measure directly the
difference between the number of vortices and antivortices, namely the {\it %
net} flux. If the simple picture of well defined phase regions with
separation of $\widehat{\xi }$ and with a choice of a minimal phase gradient
between these regions (the geodesic rule) being strictly correct, than the
r.m.s net-flux should be $\simeq n_{i}^{1/4}$. However, these arguments are
unlikely to be strictly correct, especially in a real superconductor, due to
the local gauge symmetry with the evolving magnetic field\cite
{Rudaz,Kibble2,Pogosian}. Moreover, the net-flux cannot be determined from
the simulations done until now\cite{Yates,Antunes} because these were
performed using the simple geodesic rule and fixed periodic boundary
conditions. For a plausible estimate of the net-flux one must allow some
relaxation of the geodesic rule\cite{Pogosian}. This idea was proposed for
first-order bubble collision, but it can be applied in principle to Zurek's
theory at the stage at which uncorrelated phase regions with separation of $%
\widehat{\xi }$ are considered. If, for example, we relax the geodesic rule
in such a way as to allow a variance of one random flux line in an area of $%
100$ $\widehat{\xi }^{2}$ (which may contains 100 vortices, antivortices and
homogenous sites), the r.m.s. net-flux will be determined through a random
walk count of $n_{i}/100$ flux lines. In this case, the net flux is $\simeq $
$\frac{1}{10}\sqrt{n_{i}}$ .When relaxing the geodesic rule, we can count
the net topological charge by integrating around the boundaries of the
system, only {\it after} we identify the vortices inside, as otherwise it
will not be self-consistent. Under this assumption, one sums the phase
differences around any given loop while allowing, with some probability, a
gradient of the phase between adjacent regions which is not minimal. The
total topological charge will depend on this probability, and may acquire
different values for the same initial arrangement of the regions having
different values of the order parameter. One can see that even when $n_{i}$
almost does not change (a small relaxation of the geodesic rule), the value
of the net-flux can become significant and rapidly approach it's maximal
statistical value of $\sqrt{n_{i}}$. Based on the above assumptions, we
estimate the realistic order of magnitude of the net flux line density as $%
\frac{1}{10}\sqrt{n_{i}}=10^{4}$cm$^{-2}$. Obviously, the net density of
flux is not affected by mutual annihilation, since an equal numbers of
vortices and antivortices disappear. It may decrease, however, as a result
of flux lines being expelled out of the film. We show below that under the
conditions of our experiment most of the net-flux should remain in the film.
Our setup, described below, can resolve net flux down to a limit of $20$\ $%
\phi _{0}/$cm$^{2}$. Note that this sensitivity, with the assumption about
the net flux being $\frac{1}{10}\sqrt{n_{i}}$, is sufficient to see the
effect even if $n_{i}$ is 4 orders of magnitude smaller than predicted by
Eq.(\ref{indens}). The above estimation was done for a homogeneous phase
transition. In our samples the superconducting transition is not completely
homogeneous for two reasons: temperature gradients arising during the quench
and a slightly different $T_{c}$ in different regions of the film. We
estimate the maximum temperature gradients as $\triangledown T\simeq $1K/cm.
The spread of $T_{c}$ in the different regions of the film is of the same
order. The homogeneous approximation holds if $v_{T}>\widehat{s}$ \cite
{Zurek2,Kibble3}, where $v_{T}$ is the velocity of the phase transition
front propagating across the film and $\widehat{s}$ is the characteristic
speed at which{\em \ } superconducting order parameter fluctuations
propagate at $\widehat{\varepsilon }$. This condition is identical to
imposing the demand that $\triangledown T<T_{c}\widehat{\varepsilon }/%
\widehat{\xi }.$ The estimated value of the RHS of the inequality is more
then $10^{3}$ K/cm, and we therefore conclude that the homogeneous
approximation is correct.

One more possible experimental geometry is a superconducting loop. One can
try to measure spontaneous flux generated during a rapid thermal quench of
this loop. The basic idea here is the same, namely that uncorrelated regions
will be generated during the quench, with random phases of the order
parameter in each one of them. The accumulated phase difference around the
loop will create a supercurrent and magnetic flux through the center of the
loop. If the width $d$ of superconductor forming the loop is of the order of
magnitude of $\widehat{\xi }$, the average (rms) number of flux quanta
generated in the loop is: $n_{\phi }=1/4\sqrt{L/\widehat{\xi }}$, where $L$
is the circumference of the loop\cite{Zurek2}. In our typical pattern, $%
L\simeq 20$mm, and $d\simeq 10\mu $m (somewhat bigger then $\widehat{\xi }$)
so we can substitute $d$ instead of $\widehat{\xi }$ and get $n_{\phi
}\simeq 11${\bf . }This simple argument is subject to the same uncertainties
as the ones encountered above for bulk film, due to the presence of the
magnetic field.{\bf \ }

\section{Experimental setup}

In order to measure the flux, we used a high $T_{c\text{ }}$SQUID placed
close to the superconducting film. The SQUID can detect (a) the net-flux
nucleated as the film is quenched through $T_{c}$, and (b) the field noise
caused by random hopping of these flux lines. The sign of the net flux
should be random for each individual quench. In contrast, the r.m.s. power
density of the noise spectrum is the same for vortices and antivortices, and
should characterize the {\it total} density of flux, rather than the net
density.

The essential part of the experimental setup is shown in Fig. 1. It is
basically a cryostat divided into two cells separated by a thin mylar
barrier. The SQUID and the sample are mounted on the two sides of the
barrier, facing one another. The distance between the SQUID and the
superconducting film is about 1mm. The cryostat is immersed in liquid
nitrogen at 77K. The whole system is carefully shielded from the earth's
magnetic field by several $\mu $-metal layers arranged inside a soft iron
container. The residual magnetic field in the cryostat was less then 0.2mG.
Additional small coil adjacent to the sample was used to null this field, as
well as for testing the field dependence of the results. The reason for
using two separate chambers is that in this arrangement the SQUID remains at
a constant temperature, while the sample can be heated and cooled
independently. To avoid spurious magnetic fields generated by the current
used in resistive heating, the film is heated using a focused light beam.
The light is introduced into the cryostat via a quartz rod, terminating at
about 2mm from the sample. The light illuminates the whole sample area
(10x10mm) and is confined within the plastic holder tube of the sample. In
order to achieve maximum efficiency of the heating, the superconducting film
is coated by a graphite layer which absorbs the light.

Cooling of the sample after the heating stops is via a strong thermal link
to liquid nitrogen, through helium exchange gas present in the cell. The
cooling rate through $T_{c}$ can be regulated by changing the pressure of
the gas, and hence its thermal conductivity. It takes about 1-2 sec to heat
the sample above $T_{c}$ ($\simeq $90K). Cooling begins immediately when the
light is turned off (by a shutter at the top of the cryostat). The maximum
cooling rate through $T_{c}$ is 20K/sec . Because the heating and cooling is
done mainly in perpendicular to the plane of the substrate, the temperature
of the sample is approximately the same along its lateral dimensions.
Moreover, the critical slowing down of the fluctuation in the film near $%
T_{c}$ does not affect the cooling rate. It is possible to keep the SQUID's
side of the cell in vacuum to avoid any heat leak from the sample to the
SQUID, in practice, it was not necessary.

To measure the temperature of the sample in real time during the
heating-cooling cycle we used a thin graphite strip painted on the back side
of the substrate. It's heat capacity is small, similar to that of the
superconducting film, and the changes of its resistance with temperature
enable us to measure the true temperature of the film in real time. The
temperature dependence of the resistance of the graphite was calibrated
against a diode thermometer. A typical temperature measurement during the
quench is shown in Fig. 2. During the heating period, the graphite strip may
be hotter then the film by about 1K, because it faces the light directly,
but the cooling rate is approximately the same. We usually raised the
temperature of the graphite to about 100K to ensure that all the film was
heated above $T_{c}$. We ascertained that the small measuring current
through the graphite strip did not affect the output of the SQUID, which was
the same with or without temperature measurement.

At different times through the experiment, we used two kinds of HTS
DC-SQUIDs: a commercial M-2700 unit made by Conductus, and a grain boundary
SQUID made in our lab. Both were operated in a flux locked loop (FLL). The
commercial SQUID has a magnetic field sensitivity of $2\cdot 10^{-4}$ Gauss/$%
\phi _{0}$ and an integrated pick up loop with an area close to 1cm$^{2}$.
The home made SQUID, made from a thin YBCO film on a SrTiO$_{3}$ bi-crystal
substrate, has a similar field sensitivity, however, its noise is larger by
a factor of 30 than the commercial SQUID. The advantage of using the home
made SQUID is that it can be mounted closer to the sample, as it is not
encapsulated. The distance between the home made SQUID and the YBCO film is
about 0.5mm, while for the commercial unit it is 1mm. Except for the
aforementioned S/N ratio, the results presented below were the same with
both SQUIDs.

Working with a HTS SQUID enabled us to keep the whole experiment at 77K and
to perform the fast thermal quench as described. Measurement of net flux was
done continuously during the heating and cooling cycle, but the noise
becomes well defined just close to equilibrium (i.e., at the end of the
cooling). If a net flux is generated in the film during the quench it should
be seen as an offset relative to the zero flux state prevailing when the
film is above $T_{c}$. As stated above, according to the Zurek scenario this
flux should increase in amplitude with the cooling rate while it's sign
should be random from one quench to the next. If a large number of vortices
and antivortices is preserved in the film at the end of the cooldown to 77K,
the residual flux noise should be stronger compared to a situation when the
film is relatively clean from vortices (for example, after a very slow
cooling or when the temperature is above $T_{c}$).

We now turn to estimate the sensitivity of the experiment. The coupling
ratio between the pick up loop and the SQUID is $10^{3}$ and the DC-flux
level change which one can{\normalsize \ }clearly{\normalsize \ }resolve is $%
\simeq \phi _{0}/100$. According to our measurements of the{\normalsize \ }%
screening{\normalsize \ }of the SQUID by the sample, a single vortex in the
sample induces (on average) about $0.5\phi _{0}$ into the pick up loop.
Dividing by the coupling ratio we get a net flux resolution (in the sample)
of $\sim 20$ net-$\phi _{0}$/cm$^{2}$. This measured value of the coupling
coefficient is consistent with calculations\cite{Abrikosov,Ferrari}. In Fig.
3. we show an example of a sequence in which an external field was applied
while the sample was cold. The field change picked up by the SQUID is about
half of that applied, which serves to estimate the coupling coefficient. As
the film is heated above $T_{c}$, the whole field penetrates the sample and
is picked up by the SQUID. We found the same coupling coefficient by
following the reverse procedure, namely cooling the film in a field, then
turning off the external magnetic field, and finally releasing the remanent
flux pinned in the sample by heating it above $T_{c}$. From these we
conclude that the samples have strong flux pinning and that the coupling
coefficient of flux into the SQUID is about 0.5. We used the same method to
test the superconducting continuity of the ring patterns. In addition, we
found that external magnetic fields of up to 10mG become trapped during
cooldown in the film very close to $T_{c}$. Thus, any spontaneously
generated flux remaining in the film after it is cooled to a temperature
outside the interval $\widehat{\varepsilon }$ below $T_{c}$, should survive
the cooldown to 77K. A net flux line density of $10^{4}\phi _{0}$/cm$^{2}$
is equivalent to a field of 2mG. The pinning site density in similar films
was estimated in ref.\cite{Ferrari} (and in ref. therein) as 1-6$\cdot
10^{10}$ cm$^{-2}$. Thus, we conclude that most of the net flux generated
during a quench should remain pinned in the film during the cooldown.

To estimate the expected field noise in the film, we rely on noise
measurements in YBCO thin films, done with a SQUID having field sensitivity
similar to ours\cite{Ferrari}. It was found that the noise power spectrum is
linearly proportional to the magnetic field in the film and has a $1/f$
dependence. The magnitude of the r.m.s. field noise (at 10Hz) with 10 Gauss
(10$^{8}$vortices/cm$^{2}$) was $10^{-8}-10^{-7}$ Gauss/$\sqrt{Hz}$ at 77K
(the exact number depending on the sample).{\em \ }The field noise of our
SQUID is $\simeq 3\cdot 10^{-9}$ Gauss/$\sqrt{Hz}$. Thus, we expect to
resolve a contribution to the noise in the $1/f$\ regime for a total flux
line density equal or larger than $10^{6}$\ cm$^{-2}$. This would be
possible if 10$^{-4}$ (or more) of the initial density survives. The
estimates done so far refer to unpatterned films (1cm$\times $1cm area). The
limit of sensitivity in experiments on films patterned into loops (with
diameter of $\simeq 7$mm) is $\sim 10$ $\phi _{0}$ net.

Finally, the residual magnetic field in the experimental cell (before
nulling), was found to be less then 0.2mG. Since any nulling procedure is
imperfect, a small constant field is always present in the film when cooling
through $T_{c}$. In order to evaluate the effect of any residual external
field on our data, quenches were performed under different fields in the mG
range, varying systematically both in magnitude and in the sign.

\section{Results and discussion}

In this study, we used several types of {\it c-axis} oriented epitaxial YBCO
films: (1) DC-sputtered YBCO on (100) SrTiO$_{3}$; (2) DC-sputtered YBCO on
(100)MgO; (3) Laser ablation deposited- YBCO on (100)SrTiO$_{3}$; and (4)
DC-sputtered YBCO on (001)NdGaO$_{3}$. We have also tested a sintered YBCO
ceramic sample. There are several important differences between the various
types of films. Generally, sputtered films have better crystallinity than
the ones grown by laser ablation. The degree of crystallinity is a measure
of the defect density in the film, and thus the density of the pinning
sites. Films grown on (001)NdGaO$_{3}$ are unidirectionally twinned, while
those grown on other substrates are bidirectionally twinned. In addition,
the films on (001)NdGaO$_{3}$ are lined by unidirectional nano-cracks,
perpendicular to the twins, and can be thought of as composed of
superconducting strips several $\mu $m wide, separated by the nano-cracks%
\cite{Koren}. Consequently, in those films all the flux lines are very close
to some boundary and can be therefore expelled from the film during
cooldown. Films of each type, and of varying thickness in the range
50nm-300nm, were tested both as grown ( 10$\times $10 mm$^{2}$) and after
patterning. The patterns used were: (1) discs (8mm diameter); (2) rings of
7mm diameter and 10 to 100$\mu $m width; (3) an array of 100 rings of 200$%
\mu $m diameter and 3$\mu $m width.

An experiment is taken by performing a thermal quench as described while
recording the SQUID's-FLL voltage output vs. time, or equivalently, vs.
temperature. For each sample we recorded many consecutive quenches. Typical
examples of such measurements performed on three different kinds of samples
are shown in Fig. 4, along with a reference measurement, that of a substrate
coated with graphite, but without any superconducting film. We show here two
consecutive quenches for each sample. It is clearly seen that there are
small flux jumps during the heating and cooling, but no residual flux after
the samples are cooled. These flux jumps may be different from sample to
sample. However, for any given sample the flux jumps seen here appear the
same through all the quenches, both in the sign and magnitude. Moreover,
these flux jumps are not affected by external magnetic fields\ or from the
field present at the feedback-modulation coil of the SQUID. Thus, these
jumps are definitely not related to the Kibble-Zurek mechanism. Possible
origin of this effect, which occurs near $T_{c}$, will be discussed
elsewhere. The fact that we see no residual flux at the end of the cooling
shows that within our resolution no flux lines were spontaneously nucleated
in the film during the quench. Our estimate of the predicted net spontaneous
flux line density is $\sim 10^{4}$ cm$^{-2}$. This should give a magnetic
field of $\sim $ 2 mG at the sample ($1$mG at the SQUID, which is about 5$%
\cdot $10$^{2}$ times bigger then our resolution). If spontaneous flux of
this magnitude was created, we should have seen a large signal during the
cooling (about a factor of 100 larger than the reproducible flux jumps shown
in Fig. 4). Even if the annihilation of vortices and antivortices and the
expelling of flux out of the film is very fast, we should have still seen a
transient signal of varying sign and magnitude as the film cooled through $%
T_{c}$. Such signals were not observed even at the fastest recording rate (
10$^{4}$data points/sec, corresponding to a temperature change of 2$\times $%
10$^{-5}$ in $\varepsilon $ between successive points) in the vicinity of $%
T_{c}$. At this sampling rate, the time $t_{m}$\ between successive data
points is two orders of magnitude smaller than the time $\widehat{t}=(\tau
_{0}\tau _{Q}^{\mu })^{1/(\mu +1)}$\ during which the system cools from $%
T_{c}$\ to $\widehat{\varepsilon }$\ during the quench. In our system, \ $%
\widehat{t}\simeq 10^{-2}$sec (the recording rate was limited by the minimal
integration time constant of the Flux Locked Loop). Similarly, negative
results were found also with the films patterned into rings. Finally, we did
not observe any increase in the magnitude of the noise at the end of the
cooldown following a quench.

In conclusion, in our experiments we found no evidence for spontaneous flux
line formation down to a limit of, {\it at least}, 10$^{-3}$ of the
predicted initial density. One possible reason for the negative result is
that the relation for $n_{i}$ (Eq. (\ref{indens})) is just a first
estimation and in practice it may be smaller. For example, results obtained
in experiments on $^{3}$He and $^{4}$He and also in numerical simulations
are smaller by as much as one or two orders of magnitude than predicted by
Eq. (\ref{indens}). The negative results in $^{4}$He may be a consequence of
a fast decay of vortex loops which takes place before the measurement even
begins. This scenario follows from the calculations done by Williams\cite
{Williams}. As he pointed out, a similar mechanism may also cause a decay of
vortex loops in a superconductor. In our case, vortex loop can be thought of
as a vortex-antivortex pair. Since this scenario involves only loops, no
initial imbalance between the number of vortices and antivortices is
permitted. If however, vortices and antivortices are not created in equal
numbers (as loops), then the imbalance between the number of vortices and
antivortices (the net flux) cannot decay via this mechanism. There is no
restriction within the Kibble-Zurek picture that the number of vortices and
antivortices should be equal. In our experiment, the mechanism proposed by
Williams would be reflected in the decrease of the total vortex density and
hence in the residual noise level at the end of the quench, but it will not
affect the anticipated {\it net} flux density (our main measuring
technique). The theory of Zurek implies that after the initial creation, the
density of vortices will decay as $\sim n_{i}\widehat{t}/t$ , where $t=0$\
at $T_{c}$. In the $^{4}$He experiment, the measurement began at a time
which is much bigger then $\widehat{t}$, while in our fast measurements we
took data with a time resolution $t_{m}\ll \widehat{t}$, and we covered the
whole interval from $t<0$ up to $t\gg \widehat{t}$. Thus, we should have
seen something at least during the initial stage, disrespectful of the
details of any subsequent decay. In addition to the work of Williams, a
reexamination of the Zurek scenario for the $^{4}$He case was done by Karra
and Rivers\cite{Karra}. Their calculation yields a vortex density consistent
with the negative experimental result. It is not clear what should be the
result of a similar calculation for a $3D$ high $T_{c\text{ }}$%
superconductor. A very interesting simulation of a rapid cooling of a normal
spot created inside a $2D$ superconductor (including pinning) was carried
out recently\cite{Ghinovker}. In particular, this simulation shows the great
importance of pinning in preserving a significant fraction of the vortex
population originating in the quench, essentially ad infinitum. It would be
highly desirable to extend this work to the case where the topological
charge during the quench is not conserved. Finally, we point out that the
very question of two separate superconductors having a well defined phase
difference before coming into contact is still under debate\cite{leggett}.
We hope that our results may shed some light on this interesting issue.

{\LARGE Acknowledgments}

The authors are grateful to G. Koren for preparing the samples. This work
was supported in part by the Israel Science Foundation, and by the VPR
Technion Fund for the Promotion of Research.

\begin{center}
\vspace{0in}{\LARGE Figure captions}
\end{center}

\begin{description}
\item  Fig. 1. A schematic layout of the essential part of the experimental
system. For clarity, only the edge of the mylar barrier is shown. The sample
itself is held inside a light-tight plastic holder to prevent the light from
reaching the SQUID. The assembly shown is surrounded by several mu-metal
shields.

\item  Fig. 2. A typical profile of the temperature vs. time during a
heating-cooling cycle. The inset shows an expanded view of the cooling
profile near $T_{c}$.

\item  \medskip Fig. 3. Response of the system to an application of an
external field of 0.3 mG. Initially, the film is at 77K, and after the
heating-cooling cycle it cools back to this temperature.

\item  Fig. 4. Typical time dependence of the flux picked up by the SQUID
for several samples undergoing heating-cooling cycles. The vertical line in
the middle of the figure separates two consecutive heating-cooling cycles.
The heating interval takes 1-2 sec at the beginning of each cycle. Note that
the cooling rates for each sample are different, as shown by the different
times on the respective horizontal axes.
\end{description}

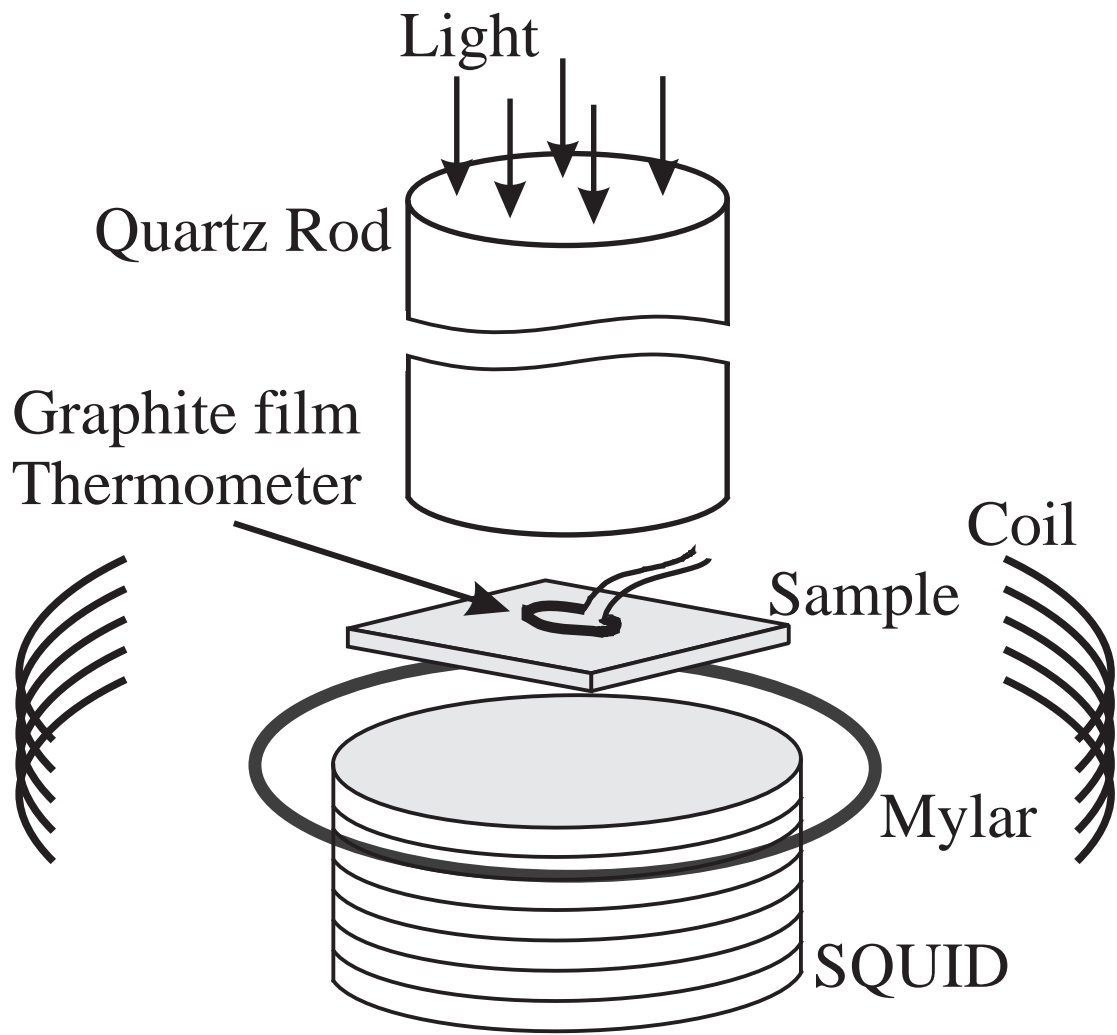
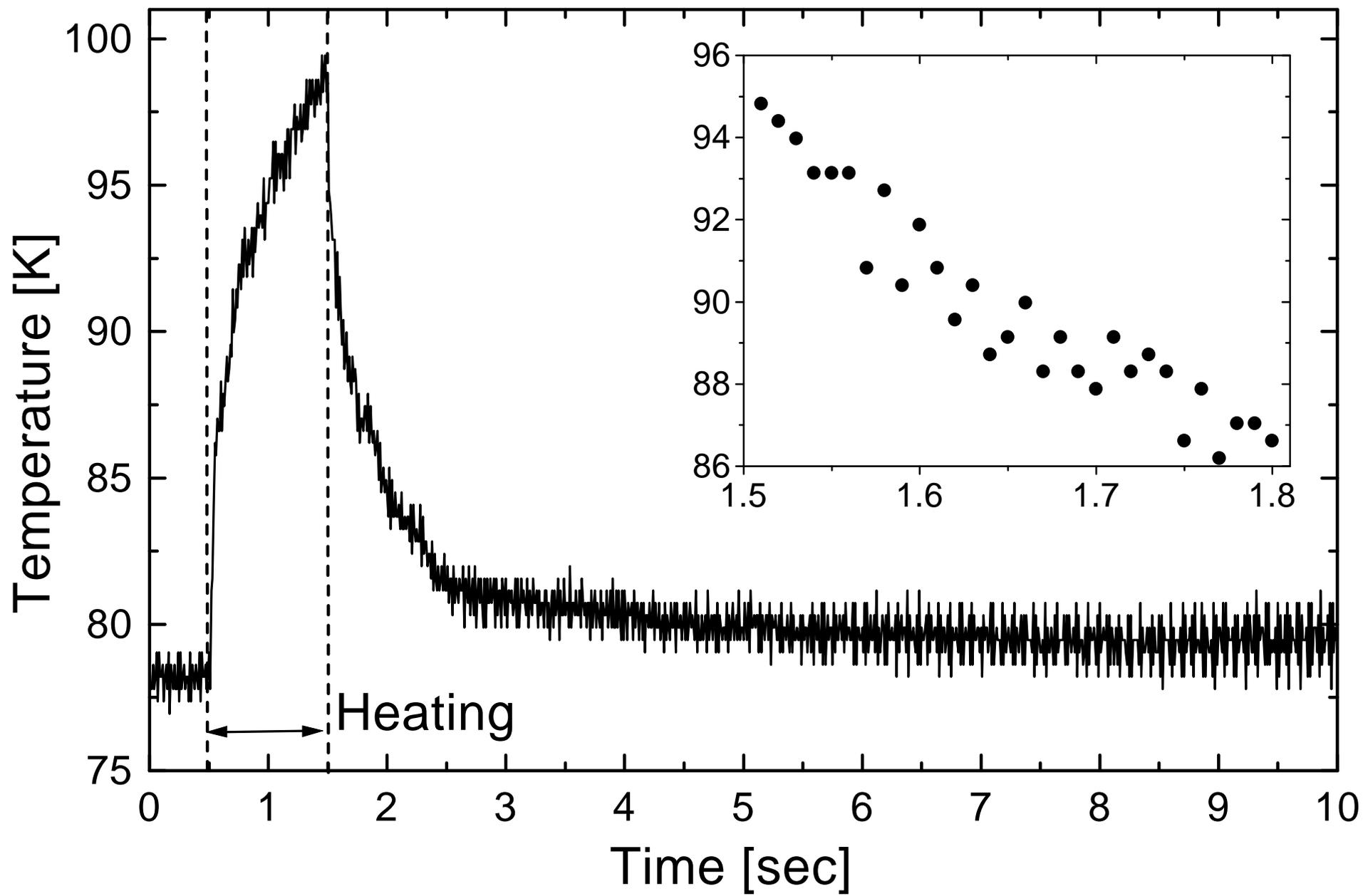
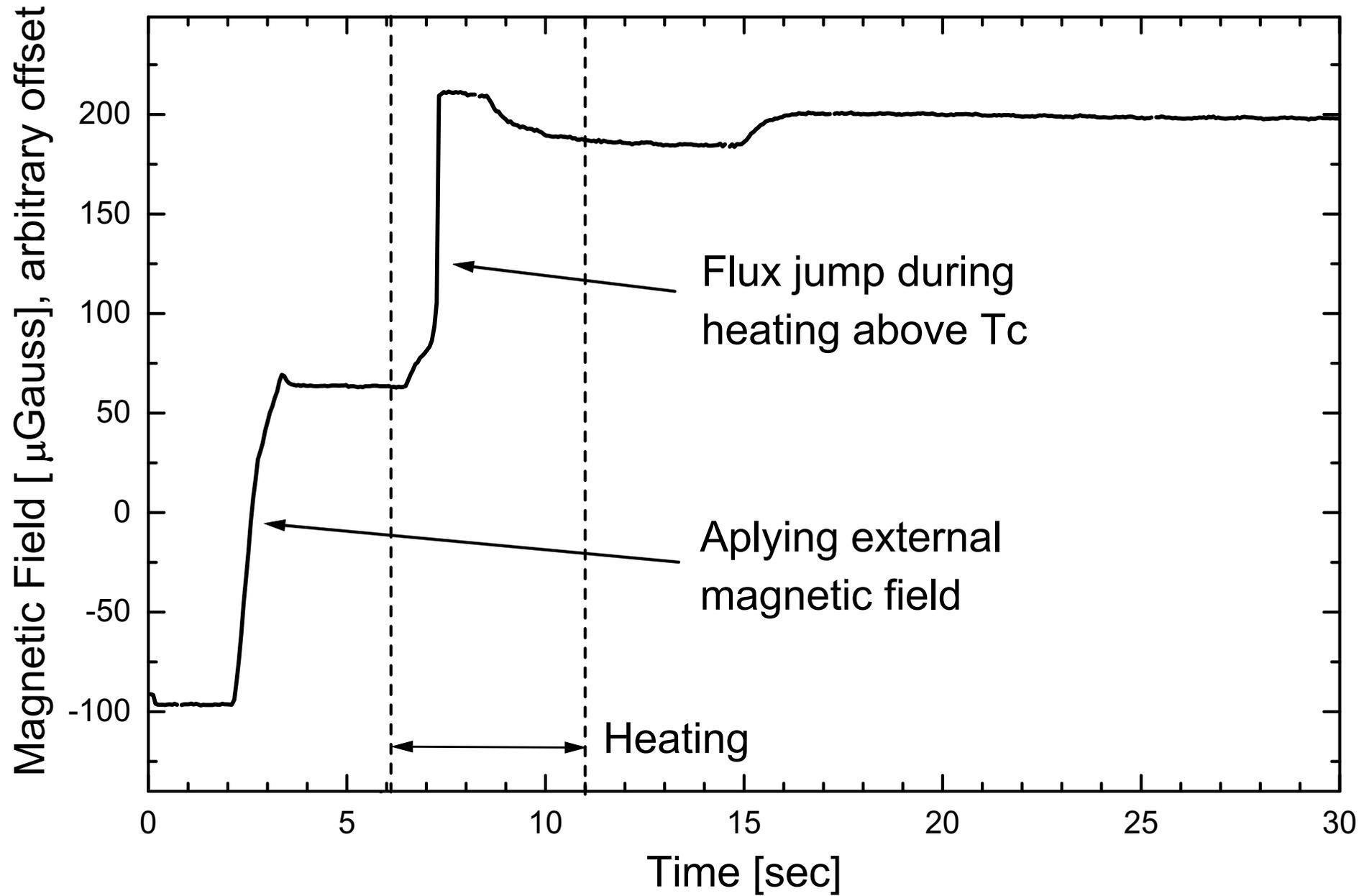
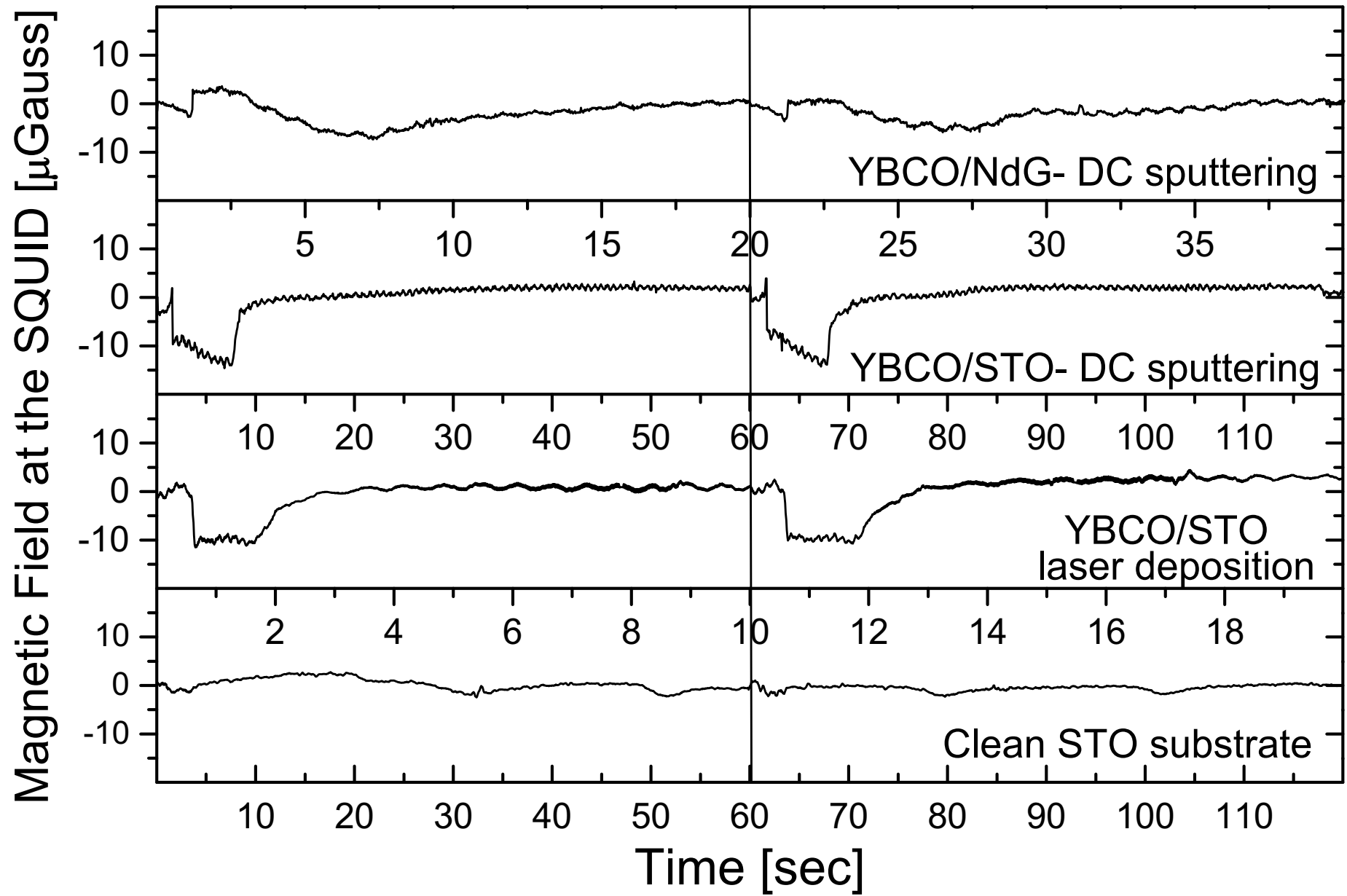
\end{document}